\documentclass[prb,twocolumn,showpacs,amsmath,amssymb]{revtex4-1}
\usepackage{graphicx} 
\usepackage{epsfig}
\usepackage{dcolumn} 
\usepackage{color}
\usepackage{changes}
\usepackage{mathtools}
\usepackage{chemformula}
\usepackage{simpler-wick}
\usepackage{siunitx}
\usepackage{esint}
\usepackage{ulem}
\usepackage{tikz-feynman}
\newcommand{\kvec}{{\bf k}}

\newcommand{\qvec}{{\bf q}} 
\newcommand{\pvec}{{\bf p}}

\newcommand{\rvec}{{\bf r}} 
 
\newcommand{\kbt}{k_{\mathrm B}T} 
\newcommand{\OO}{{\overline \Omega}}

\begin{document}

\title{Anomalous thermopower from the drag of overdamped collective modes}
\author{G. Mirarchi$^1$, G. Seibold$^2$, M. Grilli$^3$, S. Caprara$^3$}
\affiliation{$^1$Institut für Theoretische Physik und Astrophysik, Universität Würzburg, Am Hubland,
97074 Würzburg, Germany\\
  $^2$Institut f\"ur Physik, BTU Cottbus-Senftenberg - PBox 101344, D-03013 Cottbus, Germany\\
$^3$Dipartimento di Fisica, Universit\`a di 
  Roma Sapienza, piazzale Aldo Moro 5, I-00185 Roma, Italy}

\begin{abstract}
Inspired by the observation of a Seebeck coefficient ratio that exhibits 
a seemingly logarithmic divergence at low temperature in high-temperature 
superconducting cuprates, we show that a mechanisms similar to the 
standard phonon drag can give rise to anomalies in the thermopower of a 
metal, if the dragged collective mode is overdamped, with a damping
coefficient that increases with lowering the temperature. Our finding adds 
a piece to the puzzle of the strange-metal behavior observed in many 
different systems and supports our proposal that overdamped charge 
density fluctuations can be responsible of such a behavior in 
high-temperature superconducting cuprates. 
\end{abstract}
\maketitle

\section{Introduction}
\label{intro}

Despite the success of Landau’s Fermi liquid (FL) theory \cite{landau},
an increasing number of violations of this paradigm has been observed in 
the last decades. Well-known examples of non-FL behavior include 
heavy-fermion systems \cite{heavyf}, 
high-temperature superconducting cuprates \cite{cupr1,cupr2},
iron-based superconductors \cite{doiron,iron}, and organic metals\cite{doiron}. For many of 
these systems, the most prominent deviation from the FL behavior is the 
linear-in-$T$ resistivity \cite{cupr1,gurvitch,proust,varma,hussey}, with 
no change in slope from the lowest to the highest temperatures, or the 
linear-in-$\omega$ scattering rate extracted from the
optical conductivity \cite{hwang}.  
These phenomena may be accompanied by thermodynamic anomalies, such as a 
(seemingly) logarithmic divergence of the ratios $C_v/T$ and $S/T$ ($C_v$ 
is specific heat and $S$ is the Seebeck coefficient) when the temperature 
is lowered \cite{michon,michon-a,gourgout}. Henceforth, we shall adopt 
the common definition of \textit{strange metal} to refer to a system 
exhibiting these anomalies.

A phenomenological theory, based on the concept of marginal FL (MFL), was introduced to capture the strange-metal behavior of 
cuprates \cite{varma-a}. Generically, the linear-in-$T$ resistivity 
is recovered whenever a low-energy scattering mechanism extends to 
frequencies smaller than $k_\text{B}T$ and is nearly isotropic on the 
Fermi surface \cite{hlubina}. Both these features are phenomenologically incorporated in the MFL theory.

Recent progress in resonant X-ray scattering experiments 
\cite{xr1,xr2,xr3,xr4} allowed to identify new charge density collective 
modes, that coexist with the well-known charge density waves 
previously predicted \cite{fss2,andergassen,caprara} and observed in the 
underdoped regime \cite{ag}. These collective modes, dubbed 
\textit{charge density fluctuations} (CDF), can be thought of as 
\textit{aborted} charge density waves, which for some reason fail to 
establish long-range correlations and live as dynamical short range
charge density excitations. CDF are present in a much wider region of the 
phase diagram of cuprates, extending to high temperatures and dopings 
\cite{science}.  We have recently shown \cite{cdf0,cdf1,cdf2} 
that the CDF-mediated electron-electron interaction and the CDF 
contribution to the thermodynamic properties can account for both the 
linear-in-$T$ resistivity and the logarithmic divergence of the specific 
heat. Indeed, CDF are weakly momentum dependent, which allows 
CDF-mediated electron-electron scattering to be essentially isotropic 
on the Fermi surface (including Umklapp processes, required to degrade 
momentum and give rise to a non-vanishing conductivity at finite 
frequencies). Furthermore, the phenomenological assumption that the damping 
of CDF can increase (in a certain doping range) with lowering the 
temperature, allows their characteristic energy scale to decrease, while 
their correlation length stays unchanged \cite{cdf1}, producing 
a \textit{shrinking} FL \cite{grilli,preprint}, that shares some features with 
the MFL. This assumption may find a justification in two-dimensional 
systems, in terms of the coupling between CDF and electron density 
diffusive modes \cite{grilli}. The main goal of this piece of work is to 
show that the same scheme is apt to describe also the anomalous 
thermopower properties of strange metals.

Our presentation is organized as follows: in Sec.\,\ref{model}, we 
introduce our model; in Sec.\,\ref{electron} we calculate the 
electron contribution to the Seebeck coefficient; in Sec.\,\ref{CDF-drag},
we calculate the CDF drag contribution to the Seebeck coefficient; 
our results and concluding remarks are 
found in Sec.\,\ref{results}; in Appendix\,\ref{app-self}, we provide 
details about the calculation
of the electron self-energy due to CDF; in Appendix\,\ref{app-transp}, we 
give more details about the calculation of the various transport 
coefficients used in our theory; in Appendix\,\ref{app-sub}, we discuss 
in some more detail the subleading corrections to the electron 
contribution to the Seebeck coefficient; in Appendix\,\ref{app-CDF-drag}, 
we give all the details about the calclualtion of the CDF drag 
contribution to the Seebeck coefficient; in Appendix\,\ref{app-simpl}, 
we introduce a simplified model that captures our main findings, making 
their physical interpretation more transparent.

\section{The model}
\label{model}

Having in mind cuprates as a prototype strange metal, we consider a 
minimal model of electrons coupled to CDF. The latter have the twofold 
effect of dressing the electrons and contributing directly to heat 
transport. The retarded propagator of CDF with wavevector $\qvec$ 
and frequency $\omega$ is \cite{science}
\[
\mathcal{D}(\qvec,\omega)=\left[M+\bar{\nu}|\qvec-\qvec_c|^2-\mathrm i\gamma\omega-\dfrac{\omega^2}{\OO}\right]^{-1}.
\]
Here, $M$ is a characteristic low-energy scale, $\bar{\nu}$ characterizes 
the CDF dispersion law (we adopt units such that $\hbar=1$ while momenta are 
measured in units of the reciprocal of the lattice constant), $\qvec_c$ 
is the characteristic wavevector, the dimensionless parameter $\gamma$ 
characterizes damping, and $\OO$ is a high frequency cutoff. The 
dressed retarded electron Green's function is given by:
\[
G\!_R(\kvec,\omega)=\left[\omega-\xi_\kvec-\Sigma(\kvec,\omega)+\mathrm i\Gamma_0\right]^{-1},
\]
where $\xi_\kvec$ is the electron dispersion with respect to the chemical 
potential, $\Sigma(\kvec,\omega)$ is the retarded self-energy due to CDF, 
to which we add a constant term $-\mathrm i\Gamma_0$ (with $\Gamma_0>0$) 
to mimic the effect of quenched impurities. The details about the 
calculation of $\Sigma(\kvec,\omega)$ are given in the 
Appendix\,\ref{app-self}. 
Since CDF are very broad in momentum space \cite{science}, 
$\Sigma(\kvec,\omega)$ is weakly momentum-dependent \cite{cdf0,cdf1}, and we 
take $\Sigma(\kvec,\omega)=\Sigma(\omega)$. Under this assumption, 
the electron Green's function depends on $\kvec$ only through $\xi_\kvec$ 
and we shall denote it by $G(\xi_\kvec,\omega)$. 

We describe the linear-response properties within the Green-Kubo formalism, 
so the Seebeck coefficient, which is the main goal of this piece of work, 
is given by \cite{mahan}:
\begin{equation}
\label{S}
S=-\frac{1}{{\mathrm e}T}\frac{\Gamma^{12}}{\Gamma^{11}},
\end{equation}
where ${\mathrm e}\simeq+$\SI{1.609e-19}{C} is the 
elementary charge, $\Gamma^{11}$ and $\Gamma^{12}=\Gamma^{21}$ 
are defined by the relations:
\[
\begin{cases}
\langle{J^\text{el}}\rangle=-\Gamma^{11}\nabla(\mu+\text{e}\phi)-\Gamma^{12}\dfrac{\nabla{T}}{T}, \\\\
\langle{J^{Q}}\rangle=-\Gamma^{21}\nabla(\mu+\text{e}\phi)-\Gamma^{22}\dfrac{\nabla{T}}{T},
\end{cases}
\]
for the electric and heat current, $\mu$ and $\phi$ being the chemical 
and electrostatic potential, respectively. These coefficients are expressed 
in terms of the current-current response functions as (details 
are given in the Appendix\,\ref{app-transp}):
\begin{equation}
\label{Gab}
\Gamma^{\alpha\beta}=\lim_{\omega{\to}0}
\frac{\text{Im}\,\chi^{\alpha\beta}_{\qvec=\bf 0}(\omega)}{\omega}.
\end{equation}
Technically speaking, these coefficients are $2{\times}2$ matrices, but 
in the presence of tetragonal symmetry and in the absence of magnetic 
fields they reduce to scalar quantities. For this reason, we shall 
consider only one component for each current. The superscripts $\alpha$ 
and $\beta$ denote the nature of the current (1 for electric current, 2 
for heat current). The response functions 
$\chi^{\alpha\beta}_{\qvec}(\omega)$ are given by bubble-like diagrams, 
with an appropriate choice for the vertices. In the static 
(zero frequency) limit, only charged quasiparticles contribute to 
$\chi^{11}_{\qvec}(\omega)$. As CDF are neutral, their effect in $\chi^{11}_{\qvec}(\omega)$ is limited to the dressing of 
electrons. On the other hand, neutral collective modes can directly 
contribute to $\chi^{12}_{\qvec}(\omega)$ through the 
heat current channel even at $\omega=0$, in complete analogy to the 
well-known phonon drag mechanism \cite{hanna,ziman}. 
It is convenient to write 
$\Gamma^{12}=\Gamma^{12}_\mathrm{el}+\Gamma^{12}_\mathrm{drag}$, 
in order to keep the electron contribution well separated 
from that of CDF. For the calculation of $\Gamma^{11}$ we consider 
the diagram:
\begin{equation}
\label{diag1}
{j_{\kvec}^\text{el}}\quad
\feynmandiagram [baseline=(a.base), horizontal=a to b] {
a [empty dot] -- [double, with arrow=0.5, quarter left, momentum={$\kvec$, $\mathrm i(\omega_n+\Omega_\ell)$}]
b [empty dot, right=2cm of a] -- [double, with arrow=0.5, quarter left, momentum={$\kvec$, $\mathrm i\omega_n$}] a,
};
\quad{j_{\kvec}^\text{el}.}
\end{equation}
Here, the fermion lines represent the electron Green's functions 
dressed by the CDF-mediated self-energy, 
$j_{\kvec}^\text{el}=v_{\kvec,x}$ is the $x$-component of the 
quasiparticles group velocity,
$\omega_n$ is the fermion 
Matsubara frequency, while $\Omega_\ell$ is the ingoing boson 
frequency. After summing over $\omega_n$, the analytic 
continuation $\mathrm i\Omega_\ell\to\omega+\mathrm i0^+$ is applied, 
and finally the limit implied by the Eq.\,\eqref{Gab} 
is taken, yielding:
\[
\Gamma^{11}=\frac{1}{N}\sum_{\kvec,\sigma}\left({j_{\kvec}^\text{el}}\right)^2{\int_{-\infty}^{+\infty}}
\bigl[\text{Im}\,G(\xi_\kvec,\omega)\bigr]^2\left[-\frac{{\partial}f(\omega)}{\partial\omega}\right]\frac{\mathrm d\omega}{\pi},
\]
where $f(\omega){\,\coloneqq\,}(e^{\beta\omega}+1)^{-1}$ is the Fermi 
function and $\beta{\,\coloneqq\,}(k_\text{B}T)^{-1}$ is 
the inverse temperature. We point out that the vertex corrections to 
$\Gamma^{11}$ vanish in the case of a momentum-independent self-energy, 
as a consequence of the Ward identity for the 
current vertex correction \cite{ward1,ward2,metz}. 
In order to carry out the sum over $\kvec$ and $\sigma$, we introduce 
the weighted density of states:
\[
\begin{gathered}
\widetilde{N}(\xi){\,\coloneqq\,}
\frac{1}{N}\sum_{\kvec,\sigma}v_{\kvec,x}^2\delta(\xi-\xi_\kvec)
=\frac{2}{N}\sum_{\kvec}
\frac{\partial^2\xi_\kvec}{\partial{k_x^2}}\theta(\xi_\kvec-\xi),
\end{gathered}
\]
such that
\[
\Gamma^{11}=\int_{-\infty}^{+\infty}\left[\int_{-
\infty}^{+\infty}\!\widetilde{N}(\xi)\bigl[\text{Im}G(\xi,\omega)\bigr]^2
\frac{\mathrm d\xi}{\pi}\right]\left[-\frac{{\partial}f(\omega)}
{\partial\omega}\right]\mathrm d\omega.
\]
The integral in $\mathrm d\xi$ can be calculated within the Allen 
approximation \cite{allen}, yielding:
\[
\Gamma^{11}=\widetilde{N}(0)\int_{-\infty}^{+\infty}
\frac{1}{2\,\bigl[\Gamma_0-\text{Im}\,\Sigma(\omega)\bigr]}
\left[-\frac{{\partial}f(\omega)}{\partial\omega}\right]\,\mathrm d\omega,
\]
which generalizes the Drude formula for electrical conductivity. 
Indeed, within our formalism, 
the electrical conductivity is $\sigma={\mathrm e}^2\Gamma^{11}$, 
the factor $\widetilde{N}(0)$ reduces to
the ratio between the electron density and the mass, in the effective 
mass approximation, while the integral in 
$\mathrm d\omega$ represents the mean scattering time.
Note that, at zero temperature, 
$\Gamma^{11}=\widetilde{N}(0)/(2\Gamma_0)$.

\section{Electron contribution to the Seebeck coefficient}
\label{electron}

For the calculation of $\Gamma_\mathrm{el}^{12}$, we consider 
a diagram analogous to \eqref{diag1}, but with an electron heat current 
instead of a particle current:
\begin{equation}
\label{diag2}
{j_{\kvec}^{Q,\text{el}}}\quad
\feynmandiagram [baseline=(a.base), horizontal=a to b] {
a [dot] -- [double, with arrow=0.5, quarter left, momentum={$\kvec$, 
$\mathrm i(\omega_n+\Omega_\ell)$}]
b [empty dot, right=2cm of a] -- [double, with arrow=0.5, 
quarter left, momentum={$\kvec$, $\mathrm i\omega_n$}] a,
};
\quad{j_{\kvec}^\text{el}.}
\end{equation}
The question of the general definition of a microscopic heat current is 
widely debated, however it is generally agreed 
that any definition of such a current is based on energy conservation 
arguments \cite{hardy,macdowell,carbogno,helfand}. 
This approach works well for conservative systems, but may fail in 
the presence of dissipation. For the electron 
heat current we do not have this kind of problem since in this case the 
choice for the heat current $j_{\kvec}^{Q,\text{el}}$ 
simply reduces to ${\xi_\kvec}v_{\kvec,x}$ \cite{taylor,mahan}. 
The calculation of $\Gamma_\mathrm{el}^{12}$ is 
performed in full analogy with the calculation of $\Gamma^{11}$, 
yielding:
\[
\begin{gathered}
\Gamma_\mathrm{el}^{12}=\int_{-\infty}^{+\infty}\!
\left[\int_{-\infty}^{+\infty}\!\widetilde{N}(\xi)\xi
\bigl[\text{Im}\,G(\xi,\omega)\bigr]^2\frac{\mathrm d\xi}{\pi}\right]\!\!\left[-\frac{{\partial}f(\omega)}{\partial\omega}\right]\!
\mathrm d\omega\\
=\widetilde{N}(0)\int_{-\infty}^{+\infty}\frac{\omega-
\text{Re}\,\Sigma(\omega)+\text{Re}\,\Sigma(0)}{2\,\bigl[\Gamma_0-
\text{Im}\,\Sigma(\omega)\bigr]}
\left[-\frac{{\partial}f(\omega)}{\partial\omega}\right]\,\mathrm d\omega.
\end{gathered}
\]
Knowing the expressions for $\Gamma^{11}$ and $\Gamma_\mathrm{el}^{12}$, 
it is possible to use Eq.\,\eqref{S} to obtain 
the contribution to the Seebeck coefficient given by the (dressed) 
electrons alone, in the low temperature 
limit:
\begin{equation}
\label{sel}
S_\mathrm{el}{\,\coloneqq\,}
-\frac{1}{{\mathrm e}T}\frac{\Gamma_\mathrm{el}^{12}}{\Gamma^{11}}=
-T\frac{\pi^2}{3}\frac{k_\text{B}^2}{\mathrm e}
\left[-\frac{1}{2}\frac{\partial^2\text{Re}\,\Sigma(\omega)}
{\partial\omega^2}\bigg|_{\omega=0}\right].
\end{equation}
Linearity in temperature is given by the fact that, in the presence of 
a constant scattering term in the self-energy, 
$\Gamma^{11}$ tends to a constant value in the low temperature limit 
while $\Gamma_\mathrm{el}^{12}$ goes to zero as $T^2$. 
Notice that our expression for $S_\mathrm{el}$ is formally identical to 
the so-called \textit{Mott formula} for thermopower 
\cite{mott}, but in our case the quantity in brackets takes the place of 
the logarithmic derivative of the 
energy-dependent conductivity, because we made use of the Allen 
approximation.
What the two expressions have in common is that, in both cases, 
the sign of the Seebeck coefficient is associated 
with the particle-hole asymmetry of the system \cite{georges}, which 
in our model is determined solely by the self-energy. In 
Appendix\,\ref{app-sub}, for the sake of completeness, we discuss 
the subleading corrections coming from the electron self-energy, in the cases of a skewed electron
density of states.

Finally, it should be stressed that Eq.\,\eqref{sel} is valid only if 
the self-energy includes an elastic scattering 
term. If the self-energy is that of an ideal FL (with no impurity 
scattering), the expression of $S_\mathrm{el}$ 
would be formally different, but would still have an asymptotic linear
behavior as a function of the
temperature. The reason is that the asymptotic trends of 
$\Gamma^{11}$ and $\Gamma_\mathrm{el}^{12}$ at low temperature 
would be $\Gamma^{11}{\,\sim\,}T^{-2}$ and 
$\Gamma_\mathrm{el}^{12}{\,\sim\,}\text{const.}$, so the trend 
of their ratio would remain unchanged.

\section{CDF drag contribution to the Seebeck coefficient} 
\label{CDF-drag}

The Mott formula predicts a Seebeck coefficient which is linear in 
temperature, relatively small (less than 1\,$\mu$V$\cdot$K$^{-1}$ at 
$T=10$\,K) and that has the same sign as the carrier charge. 
These are not the features that are usually observed in many standard 
metals. The main reason 
for this discrepancy is that the Mott formula is valid only under the 
Bloch conditions, i.e., when the phonons are assumed to 
be at equilibrium, while at low temperatures (below the Debye temperature) 
the effects of non-equilibrium phonons become 
sizable and, in many cases, they become dominant \cite{hanna,ziman}. 
This mechanism, proposed long ago 
\cite{gurevich1,gurevich2}, is commonly known as phonon drag \cite{ziman-book}. We propose 
here a mechanism analogous to phonon drag, with 
the role of the phonons played by the CDF. Henceforth, we shall refer 
to this mechanism as \textit{CDF drag}. The diagram 
associated with the standard phonon drag at the lowest order of 
perturbation is \cite{baumann} 
\begin{equation}
\label{diag3}
\def\stack#1{\begin{subarray}{c}#1\end{subarray}}
j^{\mathrm{CDF}}_{\pvec}\;
\begin{tikzpicture}[baseline=(a.base)]
\begin{feynman}
\vertex (a) at (-2.75,0);
\vertex (c) at (0,1.5);
\vertex (b) at (2.75,0);
\vertex (d) at (0,-1.5);
\diagram*{
(a) -- [photon, quarter left, looseness=1, momentum={\(\stack{\pvec,\ \mathrm{i}(x_m+\Omega_\ell)}\)}] (c)
    -- [fermion, quarter left, looseness=1, momentum={\(\stack{\kvec,\ \mathrm{i}(\omega_n+\Omega_\ell)}\)}] (b)
    -- [fermion, quarter left, looseness=1, momentum={\(\stack{\kvec,\ \mathrm{i}\omega_n}\)}] (d)
    -- [photon, quarter left, looseness=1, momentum={\(\stack{\pvec,\ \mathrm{i}x_m}\)}] (a),
(d) -- [fermion, momentum={\(\stack{\kvec-\pvec\\ \mathrm{i}(\omega_n-x_m)}\)}] (c),
};
\fill (a) circle (2.25pt);
\filldraw[fill=white] (b) circle (2.25pt);
\end{feynman}
\end{tikzpicture}
\;
j^{\mathrm{el}}_{\kvec}
\end{equation}
and we shall use this diagram to calculate $\Gamma_\mathrm{drag}^{12}$, where $j^\text{CDF}_\kvec$ is the CDF
heat current and the wavy lines are associated to CDF propagators. After summing over the Matsubara 
frequency in the loops, we find:
\[
\Gamma_\mathrm{drag}^{12}=\frac{2g^2}{N^2}\sum_{\kvec,\pvec}j^\text{CDF}_{\pvec}j^\text{el}_{\kvec}
\bigl[f(\xi_\kvec)-f(\xi_{\kvec-\pvec})\bigr]\bigl(I_{\kvec,\pvec}^a+I_{\kvec,\pvec}^b\bigr).
\]
Further details, as well as the definitions of $I_{\kvec,\pvec}^a$ 
and $I_{\kvec,\pvec}^b$, are found in 
the Appendix\,\ref{app-CDF-drag}. Unlike the fermion case, the problem 
of defining the 
heat current for CDF is subtle, 
because of their damped character. We propose to deduce the heat current 
from the internal energy of the CDF,
which is in turn derived form the free energy of the CDF,
\[
F=\frac{1}{2\beta}
\sum_\qvec\sum_{\omega_n}\log
\left[\frac{\beta\mathcal{D}^{-1}(\omega_n,\qvec)}{\pi}\right].
\]
The sum over $\omega_n$ is formally divergent, but can be regularized 
by standard quantum 
field theory techniques. Then, the expression for the internal energy can be written as $U=\sum_\qvec U_\qvec$, where the single-mode energy contribution $U_\qvec$ is given by
\[
U_\qvec=\sum_{\eta=\pm}
\frac{\omega_\eta}{2}
\left[\frac{1}{\beta\omega_\eta}+\dfrac{\psi\Bigl(1+
\mathrm i\dfrac{{\beta}\omega_\eta}{2\pi}\Bigr)-
\log\Bigl(\mathrm i
\dfrac{{\beta}\omega_\eta}{2\pi}\Bigr)}{\mathrm i\pi}\right].
\]
Here, the quantities
\[
\begin{gathered}
\omega_\pm{\,\coloneqq\,}\pm\sqrt{\overline{\Omega}\,m_\qvec-\frac{\gamma^2\,\overline{\Omega}^2}{4}}\,-\,
\mathrm i\frac{\gamma\,\overline{\Omega}}{2}
\end{gathered}
\]
are the poles of the CDF propagator, 
$m_\qvec {\,\coloneqq\,}M+\bar{\nu}|\qvec-\qvec_c|^2$, and $\psi(z)$ is 
the digamma function. While not obvious at first glance, our expression 
for the CDF internal energy is real for all $\gamma{\,\ge\,}0$ and it 
varies continuously with $\gamma$, even when crossing the different 
damping regimes. Remarkably, in the limit $\gamma{\to}0^+$, 
$\omega_\pm=\pm(\overline{\Omega}\,m_\qvec)^{1/2}$ 
become equal and opposite real frequencies, the quantity inside the 
round brackets becomes the Bose function evaluated at those frequencies, 
and the internal energy becomes that 
of a system of standard phonons with dispersion 
$\omega_\qvec=(\overline{\Omega}\,m_\qvec)^{1/2}$. The deviation 
from the pure phonon case due to damping, within 
our description, is embodied in two distinct effects: on the one hand, 
there is the splitting of the energy 
$\omega_\qvec$ into $\omega_+$ and $\omega_-$, each of which, however, 
gives half a contribution to the thermodynamics 
of the system, on the other hand, we have a modification of the 
statistical occupation function, which would usually 
be the Bose function. This second effect is not relevant at the 
diagrammatic level, as the current entering the diagram is just a 
single-particle property, while the boson Matsubara sums provide the 
actual Bose function. Nevertheless, we take 
into account the first effect by simply considering the arithmetic 
mean of the vertices given separately by the two frequencies:
\[
j^\text{CDF}_{\qvec}=\frac{1}{2}\bigl(\omega_+\nabla\!_\qvec\omega_++\omega_-\nabla\!_\qvec\omega_-\bigl)=\frac{1}{2}\overline{\Omega}\,\nabla\!_\qvec{m_\qvec}.
\]
It is worth noting that $\gamma$ disappears, and the heat current for CDF 
is the same as in the case of undamped phonons 
with dispersion $\omega_\qvec=(\overline{\Omega}\,m_\qvec)^{1/2}$. 
In the Appendix\,\ref{app-CDF-drag}, we shall give arguments 
concerning the asymptotic behavior of $\Gamma_\mathrm{drag}^{12}$ 
at low temperature. Numerically, it is possible to show 
that, in a reasonable range of parameters and for $\gamma$ large 
enough, $\Gamma_\mathrm{drag}^{12}$ is quadratic in 
$\gamma$ and quadratic in $T$ in the low temperature limit,
$\Gamma_\text{drag}^{12}\propto{T^2\gamma^2}$.
The same asymptotic trend is clearly valid also for 
$\Gamma_\mathrm{drag}^{12}/\Gamma^{11}$, as $\Gamma^{11}$ 
tends to a finite value in the low temperature limit, which is 
independent of $\gamma$. In Appendix\,\ref{app-simpl}, we introduce
a simplified model, that allows to capture the asymptotic behavior 
of the CDF drag contribution to the Seebeck coefficient in a more 
transparent way.

It should be noted that the sign 
of $\Gamma_\mathrm{drag}^{12}$, unlike the electron case, does not 
have an immediate interpretation in terms of 
particle-hole symmetry, and it crucially depends on the model 
parameters. For the Seebeck coefficient to be positive, 
the condition $\Gamma_\mathrm{drag}^{12}<0$ must hold. 
The ratio $\Gamma_\mathrm{drag}^{12}/\Gamma^{11}$ has the dimension 
of an energy, so the constant which expresses 
linearity with $T^2\gamma^2$ must be that of $k_\text{B}^2$ divided by 
an energy. When this constant is negative, we can formally write:
\begin{equation}
\label{approx12}
\frac{\Gamma_\mathrm{drag}^{12}}{\Gamma^{11}}{\,\simeq\,}-\frac{1}{\beta^2\varepsilon(\gamma)},
\end{equation}
which defines the energy scale $\varepsilon(\gamma)$, that we consider 
to be positive and temperature-independent. Based on our analysis (see Appendix\,\ref{app-CDF-drag}), 
we can state that $\varepsilon(\gamma)$ is asymptotically proportional 
to $1/\gamma^2$. The CDF drag contribution to the Seebeck coefficient 
can be expressed as:
\[
S_\mathrm{drag}{\,\coloneqq\,}-\frac{1}{{\mathrm e}T}\frac{\Gamma_\mathrm{drag}^{12}}{\Gamma^{11}}{\,\,\simeq\,\,}
T\frac{k_\text{B}^2}{\text{e}}\frac{1}{\varepsilon(\gamma)}.
\]

\section{Results and conclusion}
\label{results}

We apply our theory to the case of cuprates. For the electron band 
dispersion we use a tight-binding model with hopping between nearest ($t=\SI{300}{meV}$), next-nearest ($t'=-\SI{82}{meV}$), and third-nearest 
($t''=\SI{17}{meV}$) neighbors, and we consider a hole doping of $24\%$. 
These parameters are suitable for the description of 
\ch{La_{1.6-x}Nd_{0.4}Sr_{x}CuO4} in the doping regime of interest. 
For the CDF, we use $M=\SI{15}{meV}$,
$\bar{\nu}=\SI{33}{meV}$ (which corresponds to $\bar{\nu}=\SI{1.3}{eV}/\text{(r.l.u.)}^2$), and $\OO=\SI{30}{meV}$. 
Finally, we take the electron-CDF coupling as
$g=\SI{320}{meV}$ and $\Gamma_0=\SI{20}{meV}$.

\begin{figure}[h]
\centering
\includegraphics[width=0.475\textwidth]{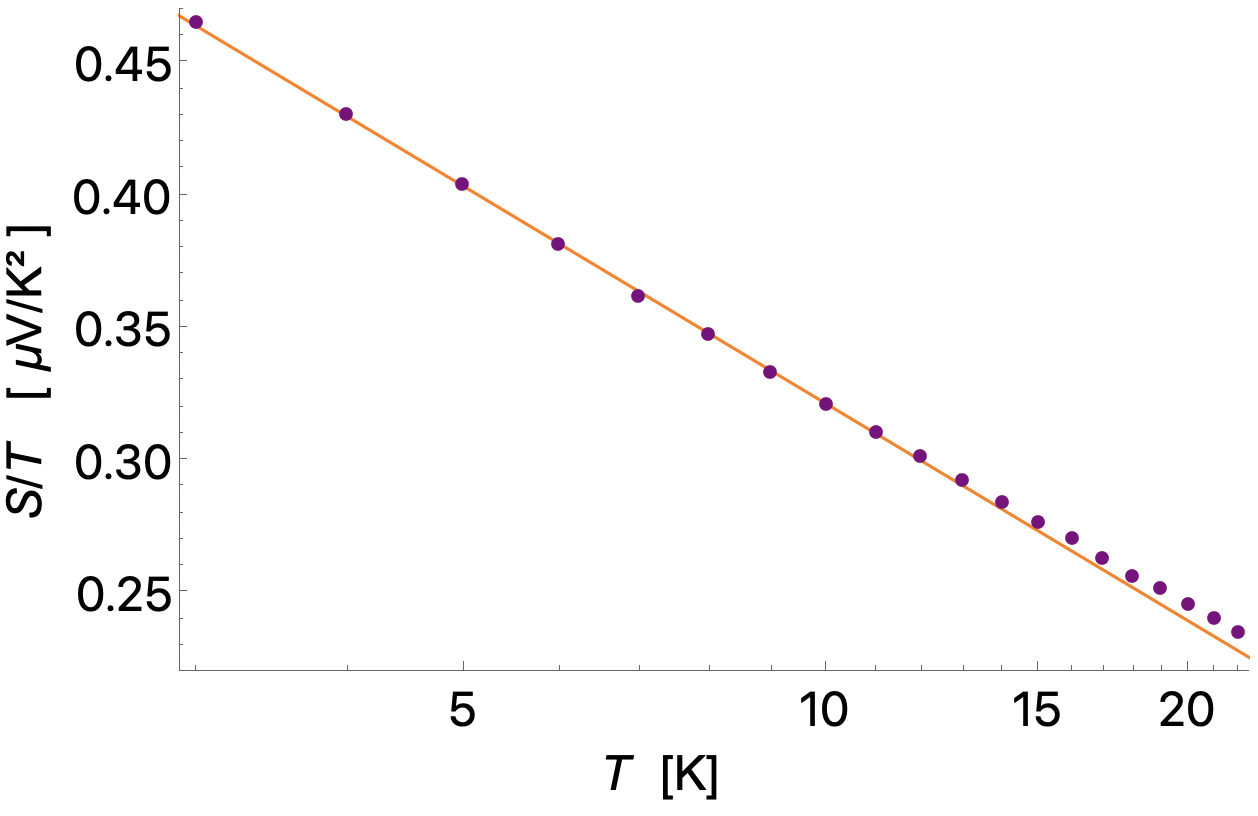}
\caption{Comparison between experimental data taken from 
Ref.\,[\onlinecite{gourgout}] (purple dots) and our theoretical calculation with the chosen parameters
(orange line), the expression we used for 
$\gamma=\gamma(T)$ is $\smash{\gamma_0\max[\log(T_0/T),1]}$, with $\gamma_0=2.66$ and $T_0=$\SI{170}{K}.}
\label{fig:fit}
\end{figure}

The only parameters that are not yet fixed are $q_c=|\qvec_c|$ and 
$\gamma$. The sign of $\Gamma_\mathrm{drag}^{12}$, 
as well as the validity of Eq.\,\eqref{approx12}, depends on $q_c$ 
through $j^\text{CDF}_{\qvec}$. Our numerical analysis 
shows that, for Eq.\,\eqref{approx12} to be valid, it is necessary 
that $q_c$ does not exceed the value of 
approximately $0.35$, roughly corresponding to $0.1$\,[r.l.u.], which is smaller than the experimentally 
reported value \cite{science}.
However, the value of $q_c$ estimated theoretically by analyzing the 
charge instability within the 
random phase approximation \cite{caprara} is also underestimated 
as compared to the experimental value. Since our theory does not include
corrections beyond the random phase approximation, one should
arguably use the corresponding value of $0.1$\,[r.l.u.], which guarantees 
the correct sign of the Seebeck effect. With $\gamma$ that grows logarithmically
from $\gamma\approx 5.5$ up to $\gamma\approx 10.5$ 
upon lowering the temperature, we are able
to reproduce the experimental data of $S/T$ for 
\ch{La_{1.6-x}Nd_{0.4}Sr_{x}CuO4} \cite{gourgout}, 
as shown in Fig.\,\ref{fig:fit}. We found that the value 
of $\varepsilon(\gamma)$ varies approximately between 
\SI{15}{meV} and \SI{45}{meV}, and is therefore comparable with the 
other energy scales of the CDF. The 
functional expression for $\gamma$ that allows us to fit the data is 
the same adopted in Ref.\,[\onlinecite{cdf1}] to fit resistivity and specific 
heat measurements, namely
$\gamma(T)=\gamma_0\max[\log(T_0/T),1]$, where $\gamma_0$ and $T_0$ 
are fitting parameters. The data in Fig.\,\ref{fig:fit}
are fitted with $\gamma_0=2.66$ and $T_0=$\SI{170}{K}. 

This results 
further supports our proposal that overdamped
CDF can be responsible for the strange-metal behavior 
of cuprates \cite{cdf0,cdf1,cdf2,preprint}. 
This finding cannot be accidental, in our opinion, strongly 
overdamped collective modes must occur in all
systems exhibiting strange-metal behavior \cite{preprint}.

As a final remark, it must be noted that, although 
$1/\varepsilon(\gamma)$ is formally 
a quadratic function of $\gamma$ in the asymptotic regime (see 
Appendices\,\ref{app-CDF-drag} and \ref{app-simpl}), it behaves
nearly linearly over the finite temperature range accessed by the 
experiment (see the argument in Appendix\,\ref{app-CDF-drag}). 
Our theory predicts that, should the experimental investigation
be extended to lower temperatures, an upward curvature
of the data would become apparent, thereby confirming their
$\gamma^2$ asymptotic behavior.
\appendix

\section{Calculation of the electron self-energy}
\label{app-self}

We compute the electron self-energy, due to the interaction with CDF, 
within the Fock approximation, which is diagrammatically represented as:

\begin{equation*}
\def\stack#1{\begin{subarray}{c}#1\end{subarray}}
\feynmandiagram [layered layout, inline=(a.base), horizontal=a to c, scale=2] {
  a -- [fermion, momentum={\(\stack{\kvec, \,\,\, \mathrm i\omega_n}\)~~~}] b 
    -- [fermion, momentum'={\(\stack{\kvec-\pvec, \,\,\, \mathrm i(\omega_n-\Omega_\ell)}\)}] c
    -- [fermion, momentum={~~~\(\stack{\kvec, \,\,\, \mathrm i\omega_n}\)}] d,
  b -- [photon, half left, looseness=1.5, momentum={\(\stack{\pvec, \,\,\, \mathrm i\Omega_\ell}\)}] c
};
\end{equation*}
The analytic expression for the imaginary part of the (retarded) 
self-energy is \cite{mazza,grilli}:
\[
\text{Im}\,\Sigma(\kvec,\omega)=\frac{g^2}{N}\sum_{\pvec}
\text{Im}\,\mathcal{D}(\kvec+\pvec,\omega-\xi_\pvec)
\bigl[f(\xi_\pvec)+b(\xi_\pvec-\omega)\bigr],
\]
where $g$ is the constant that couples electrons and CDF, with the 
dimension of an energy, $\mathcal{D}(\kvec,\omega)$ is 
the retarded CDF propagator, given in the main text, and 
$b(\omega){\,\coloneqq\,}(e^{\beta\omega}-1)^{-1}$ is the Bose 
function. Of course, $\Sigma(\kvec,\omega)$ has a negative imaginary 
part, being a retarded function. The real part can be 
obtained through a Kramers-Kronig relation:
\[
\text{Re}\,\Sigma(\kvec,\omega)=\fint_{-\infty}^{+\infty}\frac{\text{Im}\,\Sigma(\kvec,\omega')}{\omega'-\omega}
\frac{\mathrm d\omega'}{\pi}.
\]
We consider two sources of scattering for electrons: the scattering 
mediated by CDF and elastic scattering 
due to quenched impurities. The former is described by the self-energy 
we have just shown, the latter is captured by adding a 
constant term to its imaginary part:
\[
\text{Im}\,\Sigma(\kvec,\omega) \quad \longrightarrow \quad \text{Im}\,\Sigma(\kvec,\omega)-\Gamma_0,
\]
where $\Gamma_0>0$ is the scattering rate due to quenched impurities. 
Of course, the shift of 
$\text{Im}\,\Sigma(\kvec,\omega)$, being independent of frequency, 
does not affect $\text{Re}\,\Sigma(\kvec,\omega)$. 
It is worth noticing that the whole $\kvec$ dependence 
of $\Sigma(\kvec,\omega)$ lies inside the explicit momentum dependence 
of $\mathcal{D}(\kvec,\omega)$, so if the latter is momentum-independent, 
so must necessarily be the self-energy too.

\section{Transport coefficients}
\label{app-transp}

To evaluate the transport coefficients, we start from the standard 
\textit{Kubo formula} for the current-current response function \cite{kubo1,kubo2}:
\[
\chi^{\alpha\beta}_\qvec(t)=
\mathrm i\theta(t)\int\langle{[J^\alpha(\rvec,t),J^\beta({\bf 0},0)]}
\rangle\,{e^{-\mathrm i\qvec\cdot\rvec}}\,\mathrm d^3\rvec.
\]
An important consequence of this equation is the so-called 
\textit{fluctuation-dissipation theorem}, which is usually 
expressed in the following form \cite{kubo3,tong}:
\[
\text{Im}\,\chi^{\alpha\beta}_\qvec(\omega)=
\frac{1}{2}(1-e^{-\beta\omega})\langle{J^\alpha J^\beta}
\rangle_{\qvec,\omega},
\]
where $\chi^{\alpha\beta}_\qvec(\omega)$ and 
$\langle{J^\alpha J^\beta}\rangle{}_{\qvec,\omega}$ are the Fourier 
transforms of the response and of the correlation functions, respectively:
\[
\chi^{\alpha\beta}_\qvec(\omega)=\int_{-
\infty}^{+\infty}\chi^{\alpha\beta}_\qvec(t)\,e^{\mathrm i{\omega}t}\,
\mathrm dt,
\]
\[
\langle{J^\alpha J^\beta}\rangle{}_{\qvec,\omega}=\int\langle{J^\alpha(\rvec,t)
J^\beta({\bf 0},0)}\rangle\,
{e^{\mathrm i({\omega}t-\qvec\rvec)}}\,\mathrm d^3\rvec\,\mathrm dt.
\]
The standard expression for the transport coefficients within linear 
response theory at $\omega\!=\!0$ and $\qvec\!=\!\bf 0$ 
is \cite{chester,mahan}:
\[
Z^{\alpha\beta}=\frac{1}{\beta}\lim_{\epsilon{\to}0^+}
\int_0^{\infty}e^{-{\epsilon}t}\biggr(\int_0^\beta
\langle{J^\alpha(-t-i\lambda)J^\beta(0)}
\rangle{\mathrm d\lambda}\biggl)\mathrm dt,
\]
where $\langle{J^\alpha(-t-i\lambda)J^\beta(0)}\rangle$ is a 
short-hand notation for the integral of the function 
$\langle{J^\alpha(\rvec,-t-i\lambda)J^\beta({\bf 0},0)}\rangle$ 
in $\mathrm d^3\rvec$, extended over the volume of the system. 
It should be stressed that, in the case of ambiguous expressions, 
it is first necessary to fix $\qvec\!=\!\bf 0$ and possibly 
set $\omega\!=\!0$ through the limit $\omega{\to}0$, as it is customary 
for the calculation 
of transport properties \cite{kadmar}. By carrying out explicitly the 
two integrals in the expression of $Z^{\alpha\beta}$, 
and by using the fluctuation-dissipation theorem, we find:
\[
Z^{\alpha\beta}=
\frac{1}{\beta}\lim_{\omega{\to}0}
\frac{\text{Im}\,\chi^{\alpha\beta}_{\qvec=\bf 0}(\omega)}{\omega}.
\]
For practical purposes, we prefer to work with the coefficients 
$\Gamma^{\alpha\beta}$, obtained by rescaling 
$Z^{\alpha\beta}$ by the inverse temperature, 
$\Gamma^{\alpha\beta}{\,\coloneqq\,}{\beta}Z^{\alpha\beta}$. 
The electrical conductivity, the Seebeck coefficient, and the 
thermal conductivity are then given by:
\begin{equation*}
\sigma={\mathrm e}^2\Gamma^{11}, ~~ 
S=-\frac{1}{{\mathrm e}T}\frac{\Gamma^{12}}{\Gamma^{11}}, ~~
\kappa=\frac{1}{T}\left[\Gamma^{22}-
\frac{(\Gamma^{12})^2}{\Gamma^{11}}\right].
\end{equation*}

\section{Subleading contributions from the electron self-energy}
\label{app-sub}

In our theory, we considered an essentially flat electron density of
states, so that the leading electronic contribution to the Seebeck 
coefficient is given by Eq.\,\eqref{sel}. In this Appendix, we shall
show that even in the presence of a skewed electron density of
states, the contributions coming from the resulting skewed
electron self-energy would be subleading. 

Let us consider a non-constant density of states
\[
N(\xi)=\frac{A}{1+\left(\frac{\xi-\xi_\mathrm{peak}}{\Delta}\right)^2},
\]
for $\xi_\mathrm{min}\le\xi\le\xi_\mathrm{max}$, and $N(\xi)=0$ elsewhere. For a finite density of electrons, one has
$\xi_\mathrm{min}<0$ and $\xi_\mathrm{max}>0$, i.e., the Fermi level falls within the band.
In the expression for $N(\xi)$, $A$ is a normalization constant with the dimensions of 
a density of states, that we shall henceforth incorporate in the definition of the dimensionless coupling $\lambda$,
$\xi_\mathrm{peak}$ marks the position of the
peak in the density of states ($\xi_\mathrm{peak}=0$ restores ph symmetry), and $\Delta$ measures the width of the peak in the density of states.

When calculating the imaginary part of the 
retarded electron self-energy due to
the coupling with CDF, in the spirit of the Sommerfeld expansion implied by the expression
of the response functions in the Allen approximation, we can focus on 
the $T=0$ value, 
\begin{eqnarray*}
-\mathrm{Im}\,\Sigma_\mathrm{R}(\omega)&=&\frac{\lambda\gamma}{M}\int_0^\omega
\frac{\mathrm d\xi}{1+\left(\frac{\xi-\xi_\mathrm{peak}}{\Delta}\right)^2}\nonumber\\
&\times&\frac{\omega-\xi}{\left[1-\frac{(\omega-\xi)^2}{M\overline\Omega}\right]^2+\frac{\gamma^2}{M^2}(\omega-\xi)^2},\nonumber\\
~\nonumber
\end{eqnarray*}
provided $|\omega|<|\xi_\mathrm{min}|,\xi_\mathrm{max}$. Here,
$M$ is a suitable average value of $m_\qvec$. It is 
immediate to show that this is an even function of $\omega$ 
if and only if $\xi_\mathrm{peak}=0$. To extract the
asymptotic behavior, it is convenient to expand the density of states for small $\xi$, taking
\[
\frac{1}{1+\left(\frac{\xi-\xi_\mathrm{peak}}{\Delta}\right)^2}\approx\frac{\Delta^2}{\xi_\mathrm{peak}^2+\Delta^2}
\left(1+\frac{2\xi_\mathrm{peak}\xi}{\xi_\mathrm{peak}^2+\Delta^2}\right),
\]
where the second term in the brackets, enforcing skewness, vanishes for $\xi_\mathrm{peak}=0$.

The main result is that the low-energy asymptotic behavior of the odd (skewed) and even part of the imaginary part of the retarded self-energy 
is described,
respectively, by the power laws $\omega^3$ and $\omega^2$, as expected in a FL. Explicit calculation for $\omega\to 0$ gives
\[
-\mathrm{Im}\,\Sigma_\mathrm{R}(\omega)\approx\frac{\lambda\gamma\Delta^2\omega^2}{2M(\xi_\mathrm{peak}^2+\Delta^2)}
\left[1+\frac{2\xi_\mathrm{peak}\omega}{3(\xi_\mathrm{peak}^2+\Delta^2)}\right],
\]
whence it is evident that the skewed part vanishes for $\xi_\mathrm{peak}=0$. It is also evident that the skewed part is expectedly
(much) smaller than the even part, i.e., the second term in the square brackets is (much) smaller than 1. 
This condition cannot be possibly violated, because the retarded self-energy cannot change sign.
The crossover to the FL regime occurs at $\omega\approx M/\gamma$.

Since a contribution of the imaginary part of
the self-energy to the fermionic Seebeck coefficient
can only come from the skewed part \cite{georges}, 
our result shows that in our model such a contribution would be subleading with respect to
the contribution given in Eq.\,\eqref{sel}.

\section{Calculation of the CDF drag diagram} 
\label{app-CDF-drag}

The explicit expression associated to the diagram \eqref{diag3} is:
\begin{widetext}

\begin{equation}
\label{chi}
\begin{gathered}
\chi^{12}_{\qvec=\bf 0}(\mathrm i\Omega_\ell)=
\frac{g^2}{N^2}\sum_{\kvec,\sigma}\sum_{\pvec}
\biggl[\frac{1}{\beta}\sum_{ix_m}j^\text{CDF}_\pvec
\mathcal{D}\bigl(\pvec,\mathrm ix_m+\mathrm i\Omega_\ell\bigr)
\mathcal{D}\bigl(\pvec,ix_m\bigr)\Lambda\bigl(\kvec,\pvec,
\mathrm ix_m;\mathrm i\Omega_\ell\bigr)\biggr],
\end{gathered}
\end{equation}
where we have introduced the fermion vertex function associated to the loop:
\[
\Lambda\bigl(\kvec,\pvec,\mathrm ix_m;\mathrm i\Omega_\ell\bigr)
{\,\coloneqq\,}
-\frac{1}{\beta}\sum_{i\omega_n}\mathcal{G}_0\!
\bigl(\xi_\kvec,\mathrm i\omega_n\bigr)
\mathcal{G}_0\!\bigl(\xi_\kvec,\mathrm i\omega_n+
\mathrm i\Omega_\ell\bigr)\mathcal{G}_0\!
\bigl(\xi_{\kvec-\pvec},\mathrm i\omega_n-\mathrm ix_m\bigr)j^\text{el}_\kvec.
\]
\end{widetext}
All Matsubara sums can be carried out analytically, by introducing 
the spectral representation for the CDF Green's functions 
and exploiting suitable known identities for these kinds of sums. Once 
the Matsubara sums have been carried out, the 
analytic continuation $\mathrm i\Omega_\ell{\,\to\,}\omega+\mathrm i0^+$ 
is applied, and the limit $\omega{\,\to\,}0$ is 
considered in the final expression. The whole procedure is 
straightforward, however it is tedious and lengthy to write, 
therefore we only exhibit the final result, which is the one used in the 
main text:
\[
\Gamma_\mathrm{drag}^{12}=\frac{2g^2}{N^2}\sum_{\kvec,\pvec}j^\text{CDF}_{\pvec}j^\text{el}_{\kvec}
\bigl[f(\xi_\kvec)-f(\xi_{\kvec-\pvec})\bigr]\bigl(I_{\kvec,\pvec}^a+I_{\kvec,\pvec}^b\bigr).
\]
The quantities $I_{\kvec,\pvec}^a$ and $I_{\kvec,\pvec}^b$ which 
appear in this expression are defined as:

\begin{widetext}

\[
I_{\kvec,\pvec}^a{\,\coloneqq\,}-
\fint_{-\infty}^{+\infty}\frac{\dfrac{\partial}{{\partial}x}
\Bigl(\bigl[\text{Im}\,\mathcal{D}(\pvec,x)\bigr]^2
b'(x)\Bigr)}{x-(\xi_{\kvec}-\xi_{\kvec-\pvec})}\frac{\mathrm dx}{\pi}, 
~~~ I_{\kvec,\pvec}^b{\,\coloneqq\,}\frac{\dfrac{\partial}{{\partial}\xi}
\Bigl(\bigl[\text{Im}\,\mathcal{D}(\pvec,\xi)\bigr]^2
b'(\xi)\Bigr)}{\text{Im}\,\mathcal{D}(\pvec,\xi)}
\Bigg|_{\xi=\xi_{\kvec}-\xi_{\kvec-\pvec}}
\fint_{-\infty}^{+\infty}\frac{\text{Im}\,
\mathcal{D}(\pvec,x)}{x-(\xi_{\kvec}-\xi_{\kvec-\pvec})}
\frac{\mathrm dx}{\pi}.
\]

\end{widetext}

Since the most relevant momentum dependence in these two integrals comes 
from the dispersions and not from $\text{Im}\,\mathcal{D}(\qvec,\omega)$, 
we shall make the simplification of fixing the momentum 
equal to $\qvec_c$ in the CDF propagators. In this way, both integrals 
depend on momenta only through the variable 
$\Delta\xi{\,\coloneqq\,}|\xi_{\kvec}-\xi_{\kvec-\pvec}|$ (there is no 
loss of generality in considering the absolute value, 
as both integrals are even functions of 
$\xi_{\kvec}-\xi_{\kvec-\pvec}$), 
namely $I_{\kvec,\pvec}^a{\,\simeq\,}I^a(\Delta\xi)$ 
and $I_{\kvec,\pvec}^b{\,\simeq\,}I^b(\Delta\xi)$. By means of 
suitable manipulations, and by exploiting the parity properties of 
the involved functions, it is possible to express this sum in such a way 
as to make the symmetries evident:

\begin{widetext}

\begin{equation}
\label{Gdrag}
\begin{gathered}
\Gamma_\mathrm{drag}^{12}=\frac{1}{v_\text{\tiny{uc}}}
\frac{2g^2}{N^2}\sum_{\kvec,\pvec}j^\text{\!\tiny\textit{C\!D\!F}}_{\pvec}
\dfrac{j^{el}_{\kvec\!+\!\frac{\pvec}{2}}+j^{el}_{\kvec\!-\!
\frac{\pvec}{2}}}{2}\bigl[f(\xi_{\kvec\!+\!\frac{\pvec}{2}})
-f(\xi_{\kvec\!-\!\frac{\pvec}{2}})\bigr]\\
\times\bigl[I^a(\xi_{\kvec\!+\!\frac{\pvec}{2}}-\xi_{\kvec\!-\!
\frac{\pvec}{2}})+I^b(\xi_{\kvec\!+\!\frac{\pvec}{2}}-
\xi_{\kvec\!-\!\frac{\pvec}{2}})\bigr].
\end{gathered}
\end{equation}

\noindent
This quantity can be conveniently expressed as:

\begin{equation}
\label{Gint}
\Gamma_\mathrm{drag}^{12}=\iint_{\mathbb{R}^2}\varrho(x,y)
\bigl[f(x)-f(y)\bigr]\bigl[I^a(x-y)+I^b(x-y)\bigr]\,\mathrm dx\,\mathrm dy,
\end{equation}

\noindent
with the definition:

\begin{equation*}
\varrho(x,y){\,\coloneqq\,}\frac{1}{v_\text{\tiny{uc}}}
\frac{2g^2}{N^2}\sum_{\kvec,\pvec}j^\text{\!\tiny\textit{C\!D\!F}}_{\pvec}
\dfrac{j^{el}_{\kvec\!+\!\frac{\pvec}{2}}+j^{el}_{\kvec\!-\!
\frac{\pvec}{2}}}{2}\delta(\xi_{\kvec\!+\!
\frac{\pvec}{2}}-x)\delta(\xi_{\kvec\!-\!\frac{\pvec}{2}}-y),
\end{equation*}

\end{widetext}

\noindent
which evidently satisfies the relationship $\varrho(y,x)=-\varrho(x,y)$, 
which also implies $\varrho(x,x)=0$. This is a function with compact 
support, which is continuous on its support and varies on the 
characteristic energy scales of the system (both bosonic and fermionic). 
It is interesting to notice that $\varrho(x,y)$ does not depend 
on the temperature, therefore the whole temperature dependence 
of $\smash{\Gamma_\mathrm{drag}^{12}}$ comes from the remaining terms of 
integral \eqref{Gint}. Now, a first observation to make is that the 
largest contribution to that integral is given by the values of $x$ and 
$y$ such that $xy<0$ and $|x-y|{\,\gg\,}\kbt$. In fact, pairs of values 
such that $xy>0$ are given negligible weight due to the difference of 
the Fermi functions, unless they are both close to zero (in the sense 
that both $|x|<\kbt$ and $|y|<\kbt$ are valid, and therefore 
$|x-y|<\kbt$). However, if we consider a regime such that the temperature 
is clearly lower than all the other characteristic scales of the system 
(such as $m$, $\OO$ or the hopping parameters) the region 
$|x|,|y|<\kbt$ constitutes a relatively small fraction of the 
integration domain (vanishing at low temperature as $T^2$), and the 
function being integrated in this region is given little weight due to 
the term $\varrho(x,y)$, which is small when the difference $|x-y|$ is 
small.

To analyze the behavior of the CDF drag contribution to the
Seebeck coefficient, let us give the following definitions:

\begin{widetext}

\begin{equation}
\label{Gx}
G(x){\,\coloneqq\,}-\bigl[\text{Im}\mathcal{D}(\qvec,x)\bigr]^2b'(x)=\frac{\gamma^2}{\Bigl[\Bigl(m_\qvec-\dfrac{x^2}{\OO}\Bigr)^2+\gamma^2x^2\Bigr]^2}\frac{{\beta}^2x^2e^{{\beta}x}}{(e^{{\beta}x}-1)^2}\frac{1}{\beta},
\end{equation}

\begin{equation}
\label{eta}
\begin{gathered}
\eta(x){\,\coloneqq\,}\frac{\dfrac{\partial}{{\partial}x}\Bigl(\bigl[\text{Im}\mathcal{D}(\qvec,x)\bigr]^2b'(x)\Bigr)}{\text{Im}\mathcal{D}(\qvec,x)}=\frac{\gamma}{\Bigl(m_\qvec-\dfrac{x^2}{\Omega}\Bigr)^2+\gamma^2x^2}\\
\times\frac{{\beta^2x^2}e^{{\beta}x}}{(e^{{\beta}x}-1)^2}\dfrac{1}{\beta}\left[\dfrac{{\beta}x\dfrac{e^{{\beta}x}+1}{e^{{\beta}x}-1}-2}{\phantom{\dfrac{1}{2}}x^2\phantom{\dfrac{1}{2}}}+4\frac{\gamma^2-\dfrac{2}{\Omega}\Bigl(m_\qvec-\dfrac{x^2}{\Omega}\Bigr)}{\Bigl(m_\qvec-\dfrac{x^2}{\Omega}\Bigr)^2+\gamma^2x^2}\right].
\end{gathered}
\end{equation}

\end{widetext}

\noindent
These functions are regular 
even functions of $x$ at any finite temperature (the quantity inside 
square brackets in equation \eqref{eta} goes to zero as $x^2$ in the 
limit $x{\to}0$), moreover they satisfy the trivial 
identity $\eta(x)\text{Im}\mathcal{D}(\qvec,x)=-G'(x)$. By using these two 
functions, we can explicitely express the integrals $I_{\kvec,\pvec}^a$ 
and $I_{\kvec,\pvec}^a$, respectively, as:

\begin{widetext}

\begin{equation}
\label{IAint}
I_{\kvec,\pvec}^a=\int_0^{+\infty}
\frac{G\bigl(\xi_\kvec-\xi_{\kvec-\pvec}+x\bigr)
+G\bigl(\xi_\kvec-\xi_{\kvec-\pvec}-x\bigr)
-2G(\xi_\kvec-\xi_{\kvec-\pvec})}{x^2}\frac{dx}{\pi},
\end{equation}

\begin{equation*}
I_{\kvec,\pvec}^b=\eta(\xi_\kvec-\xi_{\kvec-\pvec})\,
\text{Re}\mathcal{D}(\pvec,x).
\end{equation*}

\end{widetext}

Let us begin by considering the integral $I_{\kvec,\pvec}^b$. As we 
already discussed, it makes sense to consider this function only in the 
case $\xi_\kvec\xi_{\kvec-\pvec}<0$, the terms satisfying the 
opposite relationship would make an essentially zero contribution to 
the sum \eqref{Gdrag} due to factor 
$\smash{\bigl[f(\xi_\kvec)-f(\xi_{\kvec-\pvec})\bigr]}$. 
At this point, it is necessary to consider the two cases 
$|\xi_\kvec-\xi_{\kvec-\pvec}|{\,\gg\,}\kbt$ and 
$|\xi_\kvec-\xi_{\kvec-\pvec}|<\kbt$. It is immediate to verify that 
the terms satisfying the first relation give a negligible contribution 
to the sum due to the behavior $\propto e^{-\beta|x|}$ of function 
$\eta(x)$ at large $|x|$.

By expressing $\Gamma_\mathrm{drag}^{12}$ through the integral \eqref{Gint}, 
the only regions of the domain that provide a relevant contribution 
are those such that $0<x<\kbt$ and $-\kbt<y<0$, or vice versa. In this 
case, we can expand $\varrho(x,y)$ as $\pm\,g^2(x-y)/\varepsilon_0$, 
where $\varepsilon_0$ is an energy scale which is related to the 
other characteristic scales of the system (such as $m_\pvec$ or $\OO$) 
and it is therefore much larger than $\kbt$. Since, in this 
case, $\text{Re}\mathcal{D}(x-y)$ simply goes to $1/m_\pvec$, we have:

\begin{widetext}

\begin{equation*}
\varrho(x,y)\bigl[f(x)-f(y)\bigr]I^b(x-y)\,\sim\,\dfrac{g^2}{\varepsilon_0}\frac{\gamma}{m_\pvec^3}\left(\frac{1}{6}+4\frac{\gamma^2-\dfrac{2m_\pvec}{\Omega}}{\beta^2m_\pvec^2}\right)\beta(x-y),
\end{equation*}

\end{widetext}

\noindent
where $|\beta(x-y)|\ll 1$ and $\beta m_\pvec\gg 1$. The contribution 
of this term to $\Gamma_\mathrm{drag}^{12}$ takes into account a further 
factor $T^2$, due to the fraction of the integration domain that gives
a non-negligible contribution to the entire integral (in order to obtain 
the dimensionless fraction it is necessary to divide by the square 
of a fermionic energy scale, comparable with the bandwidth). From 
these observations we conclude that the contribution 
of $\smash{I_{\kvec,\pvec}^b}$ to $\smash{\Gamma_\mathrm{drag}^{12}}$ is 
linear in $\gamma$ and vanishes at low temperature as $\smash{T^2}$.

Let us now focus on $\smash{I_{\kvec,\pvec}^a}$ and consider a regime 
in which the temperature is the lowest energy scale of the system 
(in particular, we also have $\kbt\ll m_\qvec/\gamma$). In this case, 
we can exploit the relation:

\begin{equation}
\label{ID1}
\lim_{\beta\to\infty}\dfrac{\beta^3 x^2 e^{\beta x}}{(e^{\beta x}-1)^2}=\dfrac{2\pi^2}{3}\delta(x).
\end{equation}
By comparison with the definition of $G(x)$ provided in equation 
\eqref{Gx} we get the following relation:

\begin{equation*}
\lim_{T\to0}\dfrac{G(x)}{T^2}=\dfrac{2\pi^2\gamma^2}{3m_\pvec^4}\delta(x).
\end{equation*}
which is valid for any finite value of $\gamma$. Now, if the difference $\xi_\kvec-\xi_{\kvec-\pvec}$ is noticeably larger than $\kbt$ we get the following asymptotic behavior:

\begin{equation}
\label{as1}
I_{\kvec,\pvec}^a\approx \dfrac{2\pi(\gamma \kbt)^2}{3m_\pvec^4(\xi_\kvec-\xi_{\kvec-\pvec})^2},
\end{equation}
Which shows the obvious $T^2$ behavior and is proportional to
$\gamma^2$. Anyway, we have to remind that the limit we are considering 
is that of low temperature at a \textit{fixed} $\gamma$. It is true 
that in our theory $\gamma$ varies with temperature, but this dependence 
is weak and is not capable of deviating from the $\gamma\kbt/m_\pvec{\,\ll\,}1$ regime at low temperature. For the sake of concreteness, in our calculation we considered temperatures lower than \SI{30}{K} and a $\gamma$ varying between approximately $5.5$ and $10.5$. Since $m=\SI{15}{meV}$, we are always in the $\gamma\kbt/m_\pvec{\,\lesssim\,}1$ regime and therefore the expression \eqref{as1} is valid. If we express $\gamma$ as $\overline{\gamma}+\delta\gamma$, where $\overline{\gamma}=8$ and $\delta\gamma$ is a parameter which varies approximately between $-2.5$ and $2.5$, we have:

\begin{eqnarray*}
I_{\kvec,\pvec}^a&\propto&(\gamma\kbt)^2\!=\!(\overline{\gamma}\kbt)^2\!
\left(1\!+\!\dfrac{2\,\delta\gamma}
{\overline{\gamma}}\!+\!\dfrac{\delta\gamma^2}
{\overline{\gamma}^2}\right)\nonumber\\
&\simeq&(\overline{\gamma}\kbt)^2\!\left(1\!+\!\dfrac{2\,\delta\gamma}{\overline{\gamma}}\right),\nonumber
\end{eqnarray*}

\noindent
where the term $\delta\gamma^2/\overline{\gamma}^2$ can be reasonably 
neglected, as even in the worst case it provides a deviation from the 
linear regime of less than $10\%$. This is the reason why an approximate 
linearity in $\gamma$ is observed at low temperature for $\gamma$ between 
$5.5$ and $10.5$. Of course, this is a kind of local linearity due to the 
fact that we are expanding a quadratic function far from its 
minimum, therefore there is no reason to believe that the line in 
question extrapolates to zero (and in fact it does not).

We have previously introduced the energy scale $\varepsilon(\gamma)$ through equation \eqref{approx12} as the ratio $-(\kbt)^2\Gamma^{11}/\Gamma^{12}_\mathrm{drag}$,
restricted in the $\gamma$ domain which guarantees this quantity to be positive. It therefore accounts for the contributions of both $I_{\kvec,\pvec}^a$ and $I_{\kvec,\pvec}^b$. As already 
pointed out in the main text, by varying $\gamma$ in this range of values, 
$\varepsilon(\gamma)$ varies approximately between \SI{15}{meV} and 
\SI{45}{meV}, which is a range comparable with the other bosonic 
energy scales of the model.
Although, in principle, this value is allowed to depend on temperature, our analysis confirms
that such dependence is negligible, at least within the range of our interest.

\section{A simplified model}
\label{app-simpl}

In this appendix, we shall derive the CDF drag contribution to the
Seebeck coefficient under the simplifying assumptions of a spherical
Fermi surface and a small CDF characteristic wavevector $q_c$. 

Under these assumptions, in the relevant kinematic regime, the 
fermion vertex appearing in \eqref{chi}
simplifies to \cite{spera-2026}
\begin{eqnarray*}
     \Lambda_x(\pvec,\Omega_\ell)&:=&\frac{g^2}{N}\sum_{\kvec,\sigma}\Lambda(\kvec,\pvec,\mathrm i x_m\approx 0,\mathrm i\Omega_\ell)\nonumber\\
    &\approx&\frac{C}{p_x}\left(1-\frac{|\Omega_\ell|}{\sqrt{k_F^2p_x^2+\Omega_\ell^2}}    \right),
\end{eqnarray*}
for a suitable value of the constant $C$, that is immaterial when discussing
the asymptotic behavior. Here, $k_F$ is the magnitude of the Fermi wave vector, which 
is constant for a spherical Fermi surface.

The bosonic heat current vertex is 
and is given by
\[
    j^\text{CDF}_\pvec=\omega_\qvec
     \frac{\partial}{\partial q_x} \omega_\qvec=\frac{1}{2} 
      \frac{\partial}{\partial q_x} \omega_\qvec^2=
     \OO\bar\nu(q_x-q_{c,x}),
\]
where $\omega_\qvec:=\sqrt{\OO m_\qvec}$ is the CDF dispersion, and 
$\nabla_{\bf q} \omega_{\bf q}$ is the CDF group velocity.

Thus, after analytical continuation $\mathrm i\Omega_\ell \to \omega+\mathrm i 0^+$, we find
\begin{eqnarray*}
    &&\mathrm{Im}\chi^{12}_{\qvec=\bf 0}(\omega)\propto \nonumber\\
    && \int_{-\infty}^{+\infty}\frac{\mathrm dy}{\pi} 
    \frac{\gamma y}{\overline M^2+(\gamma y)^2}
\int_{-\infty}^{+\infty}
\frac{\mathrm dz}{\pi} \frac{\gamma z}{\overline M^2+(\gamma z)^2}\nonumber \\
    && 
\times\left[b(y)-b(z)\right]\delta(y-z-\gamma \omega)\sum_\pvec
\Lambda_x(\pvec,\Omega)j^\text{CDF}_\pvec ,\nonumber 
\end{eqnarray*}
where
$\overline M$ is a suitable average of $m_\qvec$ and we considered
the limit of a large $\OO$, since the dependence of the
integrals on $\OO$ is very weak, due to their good convergence.
The dependence on $\gamma$ entirely arises from the 
integration of the two Bose propagators, while the 
contribution of the vertices is regular and yields an 
overall multiplicative factor. Taking the small $\omega$ 
limit introduces a derivative of the Bose 
function
\[
\frac{\mathrm{Im}\chi^{12}_{\qvec=\bf 0}(\omega)}{\omega}\propto 
   \int_{-\infty}^{+\infty}
   \frac{\mathrm dz}{\pi} \left[\frac{\gamma z}{\overline M^2+(\gamma z)^2}\right]^2
\left[\frac{\partial b(z)}{\partial z}\right].
\] 
After the change of variable $\zeta=\gamma z$, we find
\[
\frac{\mathrm{Im}\chi^{12}_{\qvec=\bf 0}(\omega)}{\omega}\propto 
   \int_{-\infty}^{+\infty}
   \frac{\mathrm d\zeta}{\pi} \frac{\zeta^2}{(\overline M^2+\zeta^2)^2}
\left[\frac{\partial b(\zeta/\gamma)}{\partial \zeta}\right].
\] 
This expression is evidently a function of two arguments, $\overline M$ and $\gamma\kbt$, and has two limiting 
behaviors: it is proportional to $\gamma \kbt$ when $\overline M\ll\gamma \kbt$, while 
it is proportional to $(\gamma \kbt)^2$ in the opposite limit 
$\overline M\gg\gamma \kbt$. 
The integral is well approximated by an interpolating function
\[
 \mathcal I (\overline M,\gamma \kbt):= \frac{(\gamma\kbt)^2}{\overline M^3\sqrt{\overline M^2+(\gamma\kbt)^2}},   
\]
whose low-temperature asymptotic behavior is 
$\approx 
(\gamma\kbt)^2/\overline M^4$, in agreement with the much more involved analysis
reported in Appendix\,\ref{app-CDF-drag}. 

Our result
suggests that the CDF drag contribution to the Seebeck coefficient is proportional to the CDF specific
heat ($\propto\gamma$, see Refs.\,[\onlinecite{cdf1},\onlinecite{cdf2}]) and to the characteristic
relaxation time of CDF ($\propto\gamma$), as it is expected for a 
transport coefficient.

\acknowledgments
The authors wish to express their gratitude to Riccardo Arpaia, 
Lucio Braicovich, Sergio Ciuchi, Carlo Di Castro, Giacomo Ghiringhelli, and 
Davide Valentinis, for many valuable discussions on the strange metal 
behavior.

G.S.  acknowledges support from the DFG under SE806/20-1. 
M.G. acknowledges financial support from the PNRR MUR Project No. PE0000023-NQSTI, and 
specifically the project “Topological Phases of Matter, Superconductivity, and Heterostructures” 
Partenariato Esteso 4-Spoke 5 (Grant No. PE4221852A63A88D).  
S.C. and M.G. acknowledge financial support from the Ateneo Research Projects:
“Models and theories from anomalous diffusion to strange-metal behavior”
(Grant No. RM12218162CF9D05),
“Non-conventional aspects for transport phenomena and non-equilibrium statistical mechanics” (Grant No. RM123188E830D258),
“Elementary excitations at the origin of glassy or hexatic behavior in low dimensional system, at and out of equilibrium” (Grant No. RM124190C54BE48D).
S.C. acknowledges financial support from the Ateneo Research Project:
“Interplay of phonons and charge collective excitations in cuprate high-$T_c$ superconductors”  (Grant No. RP125199B9FDBFE4).

\end{document}